\documentclass{article}
\usepackage{cite,graphicx,subfigure}
\usepackage[small,bf,up]{caption}

\usepackage{setspace}
\begin{document}
\title{Design of Plasmon Cavities for Solid-State Cavity Quantum Electrodynamics Applications}
\author{Yiyang Gong, Jelena Vu\v{c}kovi\'{c} \\
	\small\textit{Department of Electrical Engineering, Stanford University, Stanford, CA 94305}}
\maketitle
\begin{abstract}
Research on photonic cavities with low mode volume and high quality factor garners much attention because of applications ranging from optoelectronics to cavity quantum electrodynamics (QED). We propose a cavity based on surface plasmon modes confined by metallic distributed Bragg reflectors. We analyze the structure with Finite Difference Time Domain simulations and obtain modes with quality factor 1000 (including losses from metals at low temperatures), reduced mode volume relative to photonic crystal cavities, Purcell enhancements of hundreds, and even the capability of enabling cavity QED strong coupling.
\end{abstract}

The modification of the spontaneous emission (SE) rate of emitters has been a pressing topic in recent research. By enhancing or suppressing emission, we could increase the efficiencies of photon sources, reduce laser threshold, and tailor sources for cavity cavity quantum electrodynamics (QED) applications such as quantum computation and quantum communications. Some researchers have demonstrated the enhancement of emission rates (Purcell effect) in solid-state by modifying the local density of optical states with photonic crystal (PC) cavities \cite{Dirk_Cavity}. However, such works face limitations in the mode volume of the cavity. One implementation that could produce smaller mode volumes, and hence higher Purcell factors, is the use of surface plasmon (SP) modes. Already, there have been reports of enhancement of photoluminesence by coupling emitters to regions of high SP density of states (DOS) \cite{Scherer_SP}. Moreover, another group has proposed coupling emitters to metallic nanowires and nanotips to enhance Purcell factors \cite{Lukin_SP}. By employing SP cavities in solid-state, we could attempt to achieve the same or even higher SE rate enhancement as with previous designs, but with simplified fabrication.

Several authors have demonstrated decreased transmission by using periodic structures to manipulate SPs \cite{Dereux_grating,Bozhevolnyi_grating}. These experiments confirm the existence of backscattering and a plasmonic band gap in metallic gratings. In addition, other groups have demonstrated that surface plasmons interfere as normal waves and set up standing waves under certain conditions \cite{Xiang_PlasmonLens}. Given such properties, it is easy to conceive of a cavity that is the marriage of the previous two devices, a cavity that contains the electromagnetic field of the plasmon mode with metallic distributed Bragg reflector (DBR) gratings on either side of the cavity. While some plasmonic DBR cavities have been proposed in previous works\cite{Xiang_PlasmonLens, Wang_2mcav}, the designs are often impractical to fabricate.

In this letter, we propose such a metallic grating cavity. The structure is shown in Figure 1(a) and is composed of gratings with thin slices of metals on either side of an uninterrupted surface, which forms the cavity. Such a grating will open a plasmonic band gap at a frequency to be determined by the grating periodicity ($a$). The periodicity of the grating that opens a plasmonic band gap at frequency $\omega$ may be determined from the dispersion relationship of SPs at a metal-dielectric interface \cite{Raether}:
\begin{equation}
	k_{sp}=\frac{\pi}{a}=\frac{\omega}{c}\sqrt{\frac{\epsilon_{d}\epsilon_{m}}{\epsilon_{d}+\epsilon_{m}}}
	\label{SPdispersion_eq}
\end{equation}
In this work, we assume that the dielectric is GaAs, with permittivity $\epsilon_{d}=\epsilon_{GaAs}=12.25$, and the metal is silver, with permittivity estimated from the Drude model as $\epsilon_{m}=\epsilon_{\infty}-(\frac{\omega_{p}}{\omega})^2$, with $\epsilon_{\infty}=1$ and the plasmon energy of silver as $\hbar\omega_{p}=8.8$eV ($\lambda_{p}=140$nm)\cite{JV_Sim}. Setting an operation energy of $\hbar\omega=$1.2eV, we determined the grating periodicity to be $a=116$nm. Although the metal is only 30nm thick, coupled modes between the air-metal interface and GaAs-metal interfaces have a negligible impact on the dispersion relation.

\begin{figure}[hbtp]
\centering
\subfigure{\includegraphics[width=.45\textwidth]{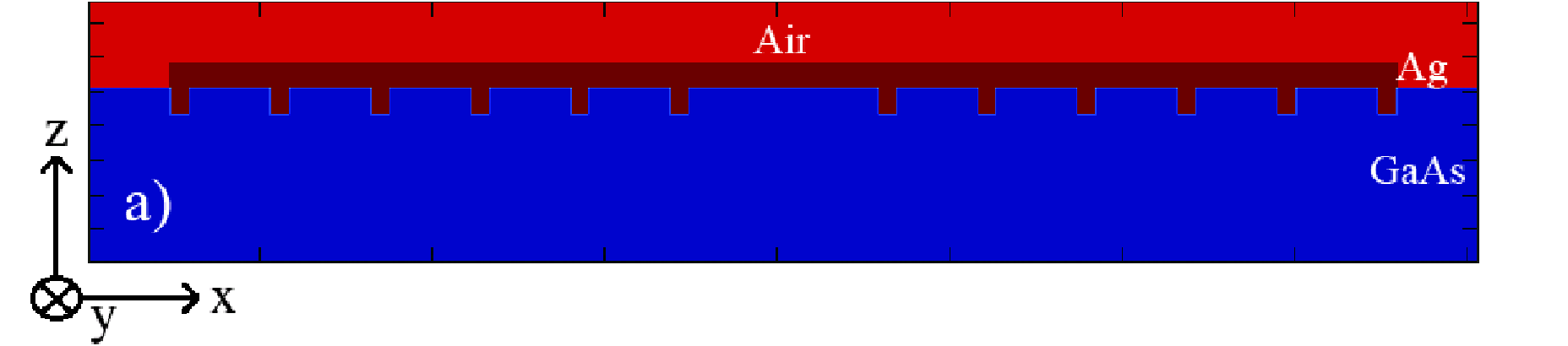}
	\label{fig:structure}}
	
\subfigure{\includegraphics[width=.4\textwidth]{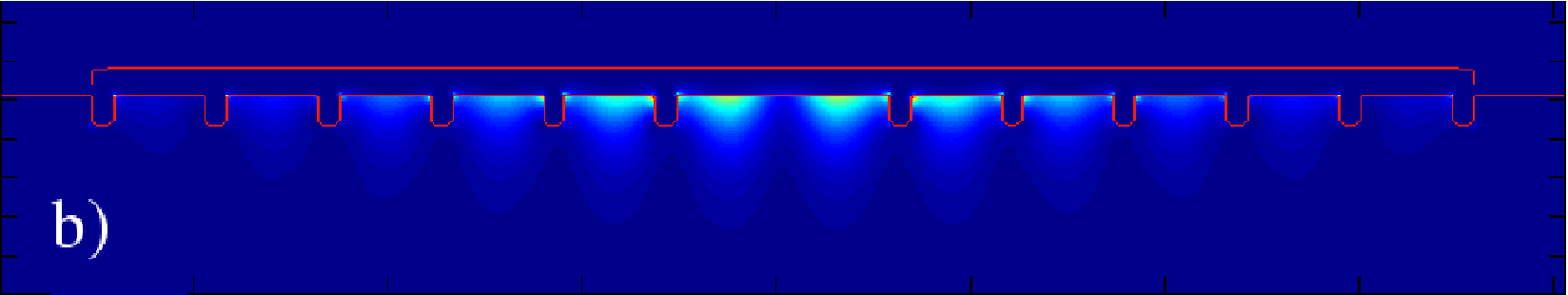}}

\subfigure{\includegraphics[width=.4\textwidth]{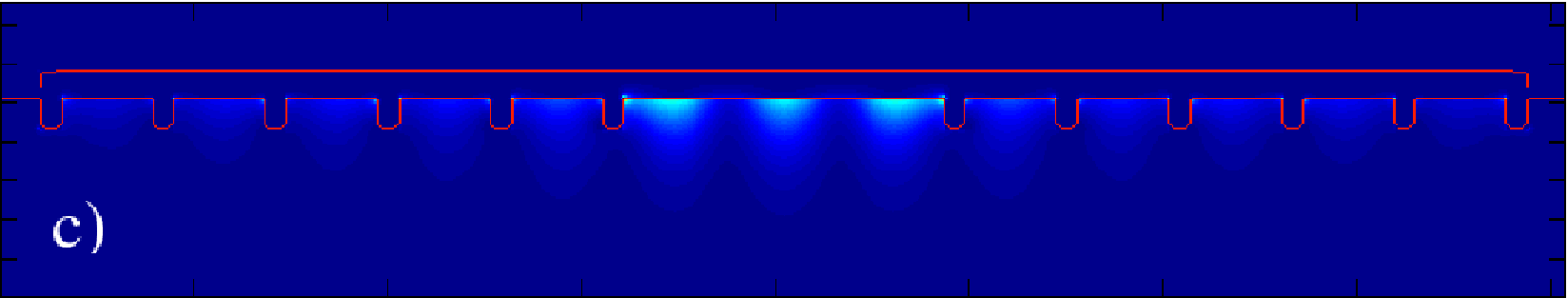}}

\subfigure{\includegraphics[width=.4\textwidth]{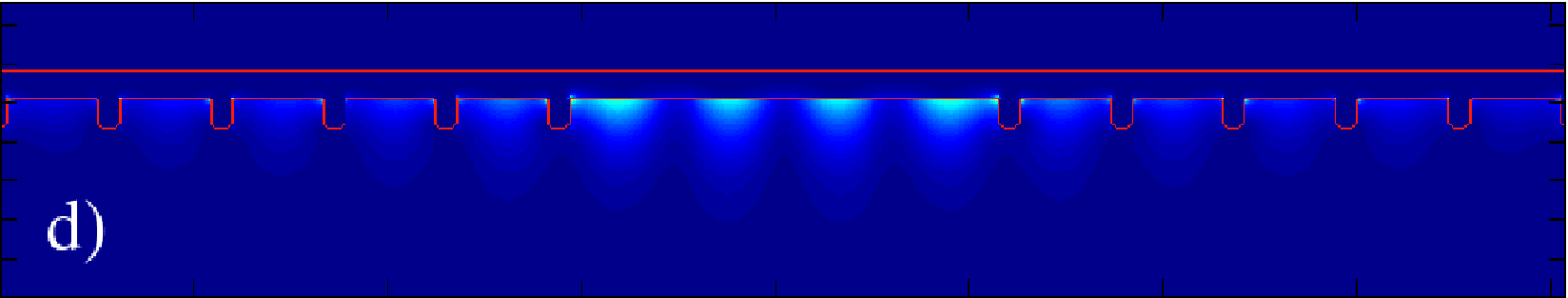}}
\caption{(a) The proposed structure. (b)-(d) Mode profiles ($|E|^2$) with cavity lengths 216nm, 328nm, and 440nm, respectively. These correspond to 2, 3, and 4 peaks of the electric field intensity inside the cavity.}
\label{fig:Modes}
\end{figure}

2D Finite Difference Time Domain (FDTD) simulations with discretization of 1 unit cell per 2nm were conducted with 5 periods of the DBR gratings on either side of a cavity and using the Drude model for the metal \cite{JV_Sim}. The depth of grooves in GaAs (filled with metal) and the metal slab layer thickness were both set at 30nm while the groove width was set at 20nm. Here, losses were also included in the Drude model with the damping energy of $\hbar\eta=2.5\times10^{-5}$eV to simulate low temperature conditions relevant for solid-state cavity QED experiments \cite{Dirk_Cavity}. This damping factor is equivalent to decreasing the non-radiative losses by approximately a factor of 2000 from their room temperature values ($\eta=\eta_{RT}/\xi$), where $\xi$ will be called the loss factor. The cavity length was then varied over multiple grating periods to determine its effect on the modes. Three prospective modes with their electric field profiles are shown in Figure 1 and the influence of cavity length on the mode quality factor (Q) and frequency is shown in Figure 2. The Q factors were calculated using $Q_{r,nr}=\omega_{0}U/P_{r,nr}$, where $\omega_{0}$ is the frequency of the mode, $U$ is the electromagnetic energy of the mode, and $P_{r,nr}$ is the power lost from the mode via radiative or non-radiative channels\cite{JV_Sim}. First, we see that indeed the modes display standing wave patterns inside the cavities. Moreover, the peak quality factors of the modes are separated by grating periods, again supporting the idea that a standing wave is formed by the reflectors on either side of the cavity. The peak quality factor is approximately 1000, and losses are dominated by radiation through the dielectric. The peaks of the quality factor all occur around $\omega=0.153\omega_{p}$, suggesting a band gap around that frequency. It is also noteworthy that the radiation parallel to the metal-dielectric interface is not the dominant pathway for losses, suggesting that increasing the number of DBRs would not enhance the overall Q or drastically change the profile of the modes. Preliminary investigations have also shown that the groove heights affect primarily the Q factors in the lateral direction, so the groove height also do not affect the overall Q. On the other hand, the duty cycle of the grating affects the SP bandgap in a non-linear manner, and thus affects the overall Q factors significantly. Finally, we see that the plasmonic modes of the DBR cavity exhibits a donor tendency, as the modes decrease in frequency as the cavity length increases.

\begin{figure}[hbtp]
\centering
\includegraphics[width=8cm]{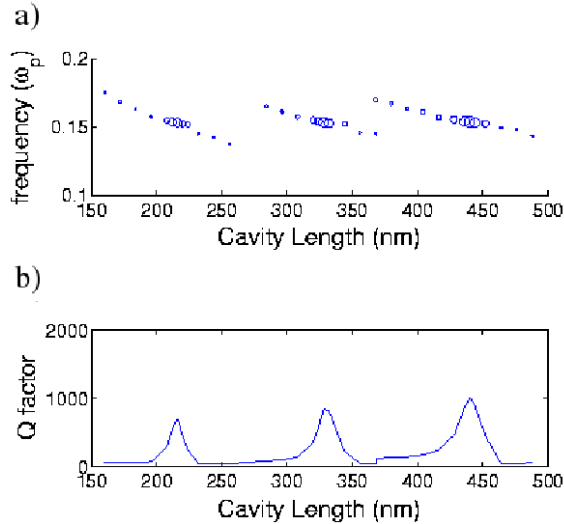}
\caption{Dependence of (a) frequency and (b) quality factor ($Q$) of the localized plasmon mode on the cavity length. In (a), the dots sizes are proportional to the mode Q factor.}
\label{fig:Q_length}
\end{figure}

Shown in Figure 3(a) is the calculated Purcell enhancement of a quantum emitter (such as an InAs/GaAs quantum dot - QD) per cavity width Y (in $\mu$m) in the y direction normal to the plane of the 2D simulation (see Figure 1). The Purcell enhancement, assuming negligible spectral detuning and non-radiative emitter decay ($\Gamma_{nr}$) is:
\begin{eqnarray}
	F&=&\frac{\Gamma_{0}+\Gamma_{nr}+\Gamma_{pl}}{\Gamma_{0}+\Gamma_{nr}} \approx 1+\frac{\Gamma_{pl}}{\Gamma_{0}}\nonumber \\
	&=&1+\frac{3}{4\pi^2}\frac{\lambda^3}{n^3}\frac{Q}{V}\left|\frac{E}{E_{max}}\right|^2
	\label{Purcell}
\end{eqnarray}
where $\Gamma_{0}$ is the emitter spontaneous emission rate in bulk, $\Gamma_{pl}$ is the emission rate of the QD coupled to the plasmon cavity mode, Q is the quality factor, and V is mode volume defined for 2D simulations as:
\begin{equation}
	V_{mode}= \frac{\int\!\!\!\int{\epsilon_{E}(x,z)|E(x,z)|^2dxdz}}{\textrm{max}\left[\epsilon_{E}(x,z)|E(x,z)|^2\right]}Y
\end{equation}
Here, Y is the width of the cavity and the effective dielectric constant is \cite{Landau_Lifshitz}:
\begin{equation}
	\epsilon_{E}(x,z)=\frac{d(\omega\epsilon(x,z))}{d\omega}= \left\{
		\begin{array}{cl}
		\epsilon_{\infty}+\left(\frac{\omega_{p}}{\omega}\right)^2, & \textrm{metal} \\
		\epsilon_{Air} \textrm{ or } \epsilon_{GaAs}, & \textrm{non-metal}
		\end{array}\right.
\end{equation}

\begin{figure}[hbtp]
\centering
\subfigure{\includegraphics[width=8cm]{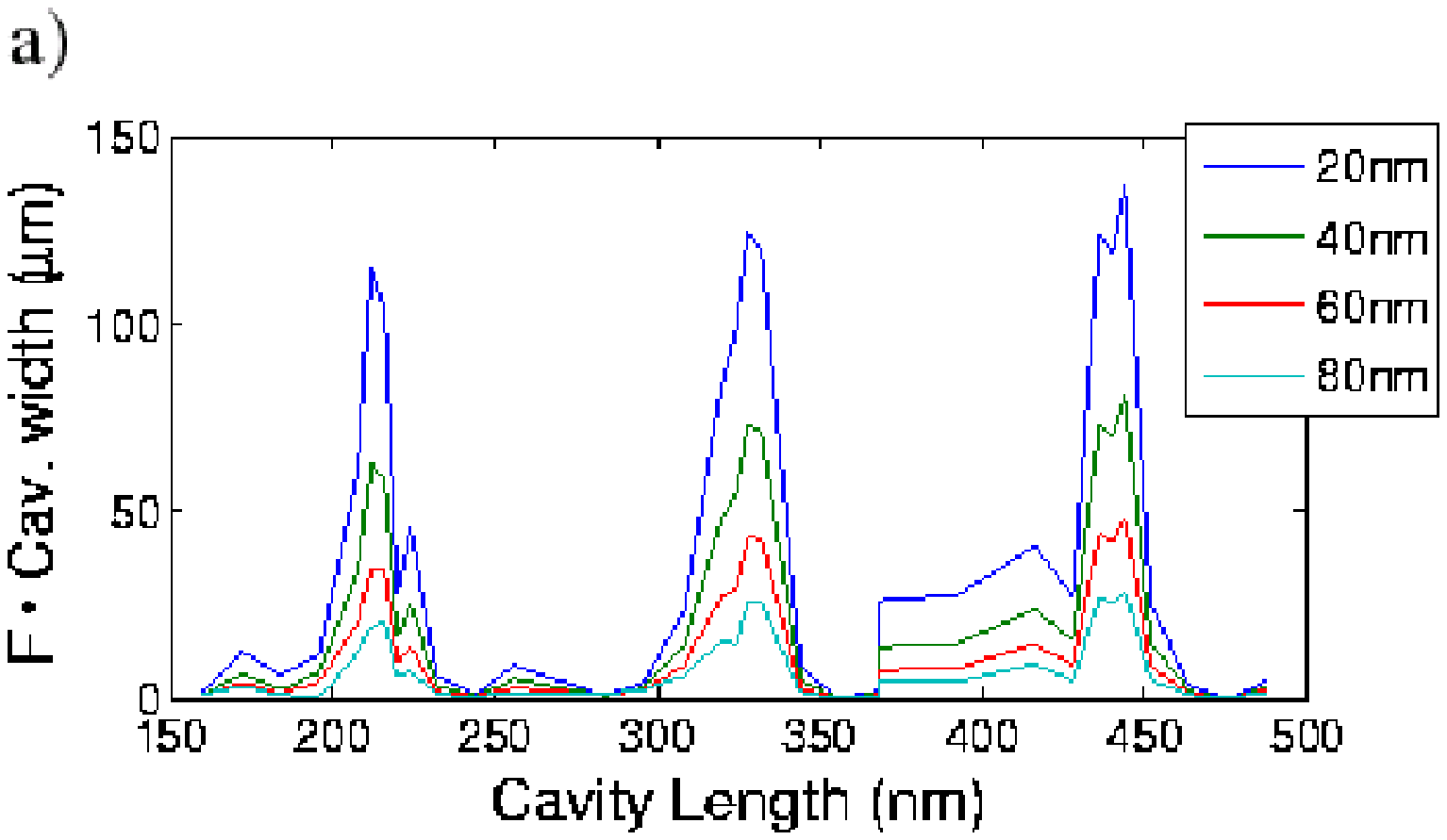}\label{fig:Purcell_length}}
\subfigure{\includegraphics[width=8cm]{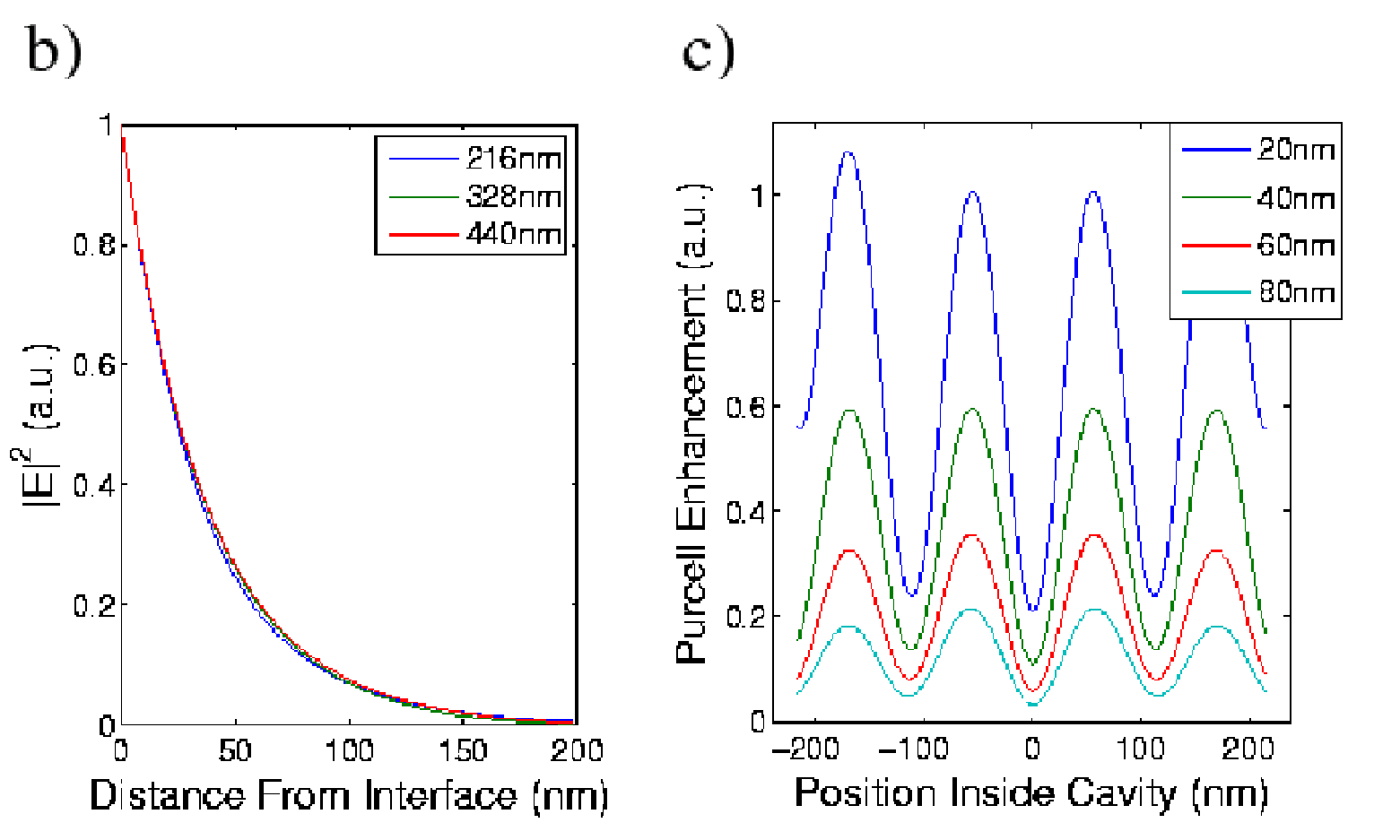}\label{fig:field_decay}}
\caption{(a) Dependence of Purcell enhancement (normalized by width of the cavity in the y direction of Figure 1) on cavity length for various emitter positions relative to the metal-dielectric interface (z direction). (b) Exponential decay of the electric field in the dielectric ($|E|^2$) away from the metal-dielectric interface, plotted for 3 different cavity lengths corresponding to maximum Purcell enhancements. The decay constant of 36nm is consistent with the plasmon modes in the band gap at the $k_{sp}=\pi/a$ point. (c) Normalized Purcell enhancement as a function of emitter position (in the x-direction) inside the 440nm cavity for four different emitter distances from the metal-dielectric interface.}
\label{fig:Purcell_all}
\end{figure}

The curves pictured in Figure 3(a) are simulated for various depths of a quantum emitter relative to the metal-dielectric interface. The larger of the cavities seem to have large tolerances of cavity lengths that could lead to the high Purcell factors. Shown in part (b) of the figure, the electric field amplitude decays exponentially away from the metal-dielectric interface, as expected for SP modes. Moreover, the decay profile of all three modes is very similar, and the decay constants of $|E|^2$ of 36nm is consistent with the plasmon mode at the $k_{sp}=\pi/a$ point. This again shows that a plasmon frequency is selectively contained by the band gap created from the gratings. Such a decay also creates a mode area of approximately (50nm)$^2$ for 2D simulations, much smaller than the $(\lambda/n)^2$ area achieved for photonic crystal cavities. If we were able to contain the field in the y direction to 50nm as well, we could in theory achieve the strong coupling regime. Namely, for an InAs/GaAs quantum dot with a dipole moment of $\mu = 10^{-28}Cm$,\cite{Lester_dipole} positioned 20nm from the metal-dielectric interface in the tail of E-field anti-node and resonant with the field, the emitter-cavity field coupling is:
\begin{equation}
	g = \mu\sqrt{\frac{\omega}{2\epsilon\hbar V_{mode}}}\left|\frac{E}{E_{max}}\right|=2\pi\times170\textrm{GHz}
\end{equation}
Note that because $E_{max}$ is not located in the middle of the cavity, but at corners of the cavities, the coupling factor is degraded from maximum values. For comparison, $\gamma$ (the emitter decay rate without a cavity, dominated by radiative decay, $\Gamma_{0}$) and $\kappa= \omega/(2Q)$ (the cavity field decay rate) are $2\pi\times1$GHz and $2\pi\times160$GHz, respectively. For such a set of parameters, $g>\gamma,\kappa$, and the onset of the strong coupling regime is reached. While degradation of Q may result from fabrication imperfections, even a ten-fold drop in Q may still result in Purcell enhancements of hundreds, enabling such a device to be used in quantum information applications\cite{Edo_EIT}.

Another interesting property of the cavity is the effective range for high Purcell effect. As shown in Figure 3(c), the electrical field amplitude follows a standing wave profile of an even mode in the x direction and exponentially decays in the z direction. While the $E_{z}$ field dominates for this mode, the contribution of the $E_{x}$ field increases as we approach the surface, making the electric field amplitude near the surface significant throughout the cavity. This effect may allow for easier coupling to quantum dots embedded throughout the substrate.

In plasmonics, one great concern is the losses of the metal at optical frequencies. However, after locating the modes with peak Purcell factors, we can increase losses by a factor of 80 and still preserve enhancement, as shown in Figure 4. Only at those lowest gains in conductivity do the radiative quality factor ($Q_{rad}$) and mode volume change (by 1\% at $\xi=25$). In the same manner, the Purcell enhancement change is dominated by the linear decrease in the absorptive Q factor ($Q_{abs}$) with loss, and only drops by a factor of 2 from the low loss ($\xi=2000$) case at $\xi=50$. This is seen if the total Q of the system is described as the parallel combination of $Q_{rad}$ and $Q_{abs}$:
\begin{equation}
	\frac{1}{Q}=\frac{1}{Q_{rad}}+\frac{1}{Q_{abs}}
\end{equation}
For loss factors greater that 100, $Q_{abs}$ is far greater than $Q_{rad}$, and thus is negligible, and the mode is not significantly perturbed. However for loss factors less than 25, the mode is changed, resulting in a diminished Q and negligible Purcell enhancement. This suggests that the sample temperature should be reduced to at least 40K (where $\xi\approx25$, \cite{Matula_Aglosses}) in order for the radiative losses to be dominant (Figure 4(b)). Such operational temperatures are already being used for solid-state cavity QED experiments \cite{Dirk_Cavity}.

\begin{figure}[hbtp]
\centering
\includegraphics[width=8cm]{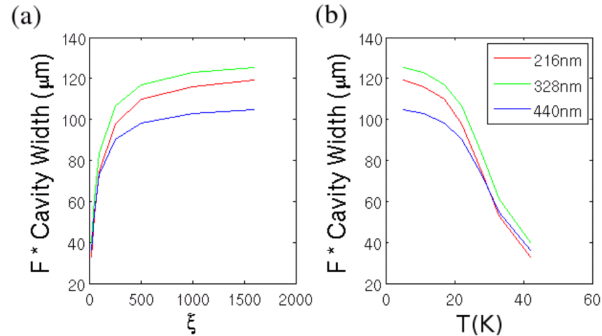}
\caption{(a) Dependence of Purcell enhancement on the loss factor, $\xi$, in the Drude model ($\xi$ is inversely proportional to the damping frequency). The Purcell enhancements are calculated for emitters 20nm from the metal-dielectric interface for three different cavity lengths. (b) Dependence of the Purcell enhancement on the temperature, which is obtained by translating the $\xi$ factor into temperatures for a residual resistivity of 1467 for silver \cite{Matula_Aglosses}.}
\label{fig:Purcell_loss}
\end{figure}

In conclusion, we have proposed and studied a plasmon cavity that could enable high quality factor ($Q=1000$), large Purcell enhancements ($F=100$), and even the strong coupling regime of cavity QED. Such devices benefit from the small mode volumes inherent with SP modes. We have also demonstrated that such a cavity indeed arises from the bandgap effect of plasmonic gratings, and that losses due to the metal do not seriously affect the operation of the device. Numerous optoelectronic and quantum optical devices could benefit from the proposed cavity design; the plasmonic cavity could enhance the out-coupling efficiency of QDs embedded in the dielectric, improving the operation of single photon emitters. Future designs would focus on containing the electromagnetic field in the third dimension, so as to fully realize the small mode volumes of the SP modes. 
 
Financial support was provided by the MURI Center for Photonic Quantum Information Systems (ARO/DTO Program No. DAAD19-03-1-0199), NSF Grant No. CCF-0507295, and the NDSEG Fellowship.

\end{document}